# Probable Decay Modes at Limits of Nuclear Stability of the Superheavy Nuclei

**M. Bhuyan**[*]

*Instituto Tecnológico de Aeronáutica, São José dos Campos, São Paulo, Brazil*

**Abstract:** The modes of decay for the even–even isotopes of superheavy nuclei of $Z$ = 118 and 120 with neutron number $160 \leq N \leq 204$ are investigated in the framework of the axially deformed relativistic mean field model. The asymmetry parameter $\eta$ and the relative neutron–proton asymmetry of the surface to the center ($R_\eta$) are estimated from the ground state density distributions of the nucleus. We analyze the resulting asymmetry parameter $\eta$ and the relative neutron–proton asymmetry $R_\eta$ of the density play a crucial role in the mode(s) of decay and its half-life. Moreover, the excess neutron richness on the surface, facets a superheavy nucleus for $\beta^-$ decays.

_______
[*] E-mail: bhuyan@ita.br

# 1. INTRODUCTION

Over the last three decades, the synthesis of superheavy nuclei has been dramatically rejuvenated owing to the emergence of the cold fusion reactions, performed mainly at GSI, Dramstart [1–6], and the hot fusion and/or the actinide based fusion reactions performed mainly at JINR, Dubna [7–12]. Through these advancements of stable nuclear beam technology, it is not only possible to synthesize superheavy nuclei but also provide impressive prospects for understanding the nuclear properties of these nuclei [1–13]. At present, the question of the mode of decay and the stability of these newly synthesized nuclei arises. While reviewing the production and decay properties of nuclei with atomic number $100 \leq Z \leq 118$, it can be seen that the sustainability of these superheavy nuclei is controlled mainly by the spontaneous fission and $\alpha$ decay processes [1–5, 7–13]. The key reason for the decay process of superheavy nuclei is the shell effect. It supplies the extra binding energy and increases the barrier height of fission [14–19]. The situation in the case of spontaneous fission is very complex as compared to the $\alpha$ decay process along the stability line of superheavy region. Further, there are other possible $\beta^-$ decay modes for a superheavy nucleus, which proceeded via the weak interaction. This process is slow and less favored as compared to spontaneous fission and $\alpha$ decay in the valley of stability.

The most stable superheavy nuclei are predicted to be located along the neutron-rich region of the $\beta$-stability line. It is not possible to reach those directly by the above mentioned fusion reactions with stable ion beams. In fact, the predicted magic proton numbers for the superheavy region are quite different within various theoretical approaches. For example, the magic proton number $Z$ = 114 was predicted in the earliest macro-microscopic calculations [20–23] and later confirmed by [15, 24]. Fully microscopic approaches predict the proton shell closure at $Z$ = 120 [25–28], and/or $Z$ =126 [29] using the chosen nucleon–nucleon interaction in mean field models. The neutron magic number $N$ = 184 is almost firmly predicted by different theoretical models [24, 26, 28]. For further experimental study of the superheavy nuclei, especially near the neutron-rich side of the nuclear chart, basic ideas of the internal structure and reaction mechanism of those nuclei from advanced theoretical approaches are required. In other words, in order to produce superheavy nuclei in the laboratory, one needs to know the internal configuration and the radioactive decay properties theoretically. Hence, the knowledge of the modes' decay and half-lives of a nucleus over a very wide range of neutron–proton asymmetry within advanced theories are essential for their synthesis process and further progress in experiments.

In this regard, we investigate different possible radioactive decay modes for the neutron rich superheavy nuclei. We have used the well-defined relativistic mean field (RMF) formalism [30–32] with the recently developed NL3* force parameter [33] for the present analysis. The model have been successfully applied in the description of nuclear structure phenomena both in $\beta$-stable and $\beta$-unstable regions throughout the nuclear landscape including superheavy nuclei [30–45]. The aim of the present study is to determine the properties of the modes of decay of neutron-rich superheavy nuclei, which may help us to answer some important open questions: (1) How far may we still move in synthesis of superheavy elements by the fusion reactions? (2) Where is the island of stability centered? (3) What are the properties of the most stable superheavy nuclei? and (4) how can one reach this region? Further, the decay properties also play a crucial role in the study of the r- process of nucleosynthesis as well as the formation of heavy and superheavy nuclei in nature [46, 47]. Here we have considered the isotopic chains $Z = 118$ and 120 with $160 \leq N \leq 204$, predicted to be the next magic valley [27, 28, 48] after $^{208}$Pb. The basic concept is that the decay process is highly influenced by the internal configuration (i.e. the arrangement of the nucleons) of the nucleus. In other words, the internal arrangement of nucleons determines the stability and modes of decay of the nucleus.

The paper is organized as follows: Section 2 gives a brief description of the relativistic mean-field formalism. The calculated results are discussed in Section 3. Section 4 includes a short summary along with a few concluding remarks.

## 2. RELATIVISTIC MEAN FIELD FORMALISM

The microscopic self-consistent mean-field calculations are the standard tool for the investigation of nuclear structure phenomena. Relativistic mean field (RMF) is one of the most popular and widely used formalism among them. It starts with the basic Lagrangian that describes nucleons as Dirac spinors interacting through different meson fields. The original Lagrangian of Walecka has taken several modifications to take care of various limitations and the recent successful relativistic Lagrangian density for nucleon–meson many-body systems [30–45] is expressed as:

$$\mathcal{L} = \bar{\psi}_i \left\{ i\,\gamma^\mu \partial_\mu - M \right\} \psi_i \; + \frac{1}{2} \partial^\mu \sigma \; \partial_\mu \sigma - \frac{1}{2} m_\sigma^2 \; \sigma^2 \qquad (1)$$

$$- \frac{1}{3} g_2 \; \sigma^3 - \frac{1}{4} g_3 \; \sigma^4 - g_s \bar{\psi}_i \; \psi_i \sigma - \frac{1}{4} \Omega^{\mu\vartheta} \; \Omega_{\mu\vartheta}$$

$$+ \frac{1}{2} m_\omega^2 V^\mu V_\mu \; - \; g_\omega \bar{\psi}_i \gamma^\mu \psi_i V_\mu - \frac{1}{4} \mathbf{B}^{\mu\vartheta} \cdot \mathbf{B}_{\mu\vartheta}$$

$$- \frac{1}{2} m_\rho^2 \mathbf{R}^\mu \cdot \mathbf{R}_\mu - g_\rho \bar{\psi}_i \; \gamma^\mu \; \boldsymbol{\tau} \; \psi_i \cdot \mathbf{R}_\mu$$

$$- \frac{1}{4} F^{\mu\vartheta} F_{\mu\vartheta} - e \; \bar{\psi}_i \; \gamma^\mu \; \frac{(1 - \tau_{3i})}{2} \; \psi_i A_\mu \, .$$

From the above Lagrangian we obtain the field equations for the nucleons and mesons. These equations are solved by expanding the upper and lower components of the Dirac spinors and the boson fields in an axially deformed harmonic oscillator basis, with an initial deformation $\beta_0$. The set of coupled equations are solved numerically by a self-consistent iteration method [31–36]. The center-of-mass motion energy correction is estimated by the usual harmonic oscillator formula $E_{\text{c.m.}} = \frac{3}{4}(41A^{1/3})$. The total quadrupole deformation parameter $\beta_2$ is evaluated from the resulting proton and neutron quadrupole moments, as

$$Q = Q_p + Q_n = \sqrt{\frac{16\pi}{5}} \left( \frac{3}{4\pi} AR^2 \beta_2 \right). \qquad (2)$$

The root-mean-square (rms) matter radius is defined as

$$\langle r_m^2 \rangle = \frac{1}{A} \int \rho(r_\perp, z) \; r^2 d\tau, \qquad (3)$$

where $A$ is the mass number, and $\rho$ is the axially deformed density. The constant strength scheme is adopted to take into account pairing correlations [35, 49–50] and evaluate the pairing gaps for neutron and proton using the BCS equations [51]. The total binding energy and other observables are also obtained by using the standard relations, given in [32]. In order to take care of the pairing effects in the present study, we have used the constant gaps for proton and neutron, as given in [35, 49, 50], which are valid for nuclei both on and away from the stability line (more details, see, e.g., [35]). We have used the recently developed NL3* [33] force parameters in the present calculation, which are able to reproduce the properties of the stable nuclei as well as the nuclei away from the $\beta$-stability line. We obtain different potentials, densities, single-particle energy, nuclear radii, deformation parameter and binding energies. For a given nucleus, there are few solutions. Among them, the solution corresponding to maximum binding energy treated as ground state for a given nucleus and other solutions are the intrinsic excited states.

## 3. RESULTS AND DISCUSSIONS

Based on the fundamental and well-established concept, we have tried to explain the modes of decay of superheavy nuclei, by means of their internal structure and sub-structure. To know the proper internal configuration, it is important to know the ground and first intrinsic excited state properties of the nucleus. The bulk properties such as binding energy (BE), root-mean-square charge radius $r_{\mathrm{ch}}$, matter radius, and the energy difference between the ground state and the intrinsic first excited state $\Delta E$ are calculated using RMF (NL3*) force parameter. The results for the isotopic chains of $Z = 118$ and 120 are listed in Tables 1 and 2, respectively. The quantity ($R_\eta$) in the last column of both the tables will be discussed in subsequent sections. From both the tables, one can point out that the ground state solutions for the isotopic chain of $Z = 118$ and 120 are deformed prolate configuration which rather follows a spherical excited solution. Further, the re-normalized nucleon numbers reflect on the total density distributions of the nucleus for a specific solution. Before going to the axially deformed density distributions, we will show a typical example of calculated spherical for $^{304}120$ in Fig. 1. Here, $\rho_p$ and $\rho_n$ are for proton and neutron density respectively, as a function of radius. From the figure, it is clear that the density of both neutron and proton uniformly spread from the center to a certain distance ($\sim r = 6$ fm), considered as the central region and the area cover from the falling point of density to the surface is taken as the surface region in the present analysis. We found some humps appear in the central region of the density, which shows the well-known shell structure of the nucleus (see [28] for more detail of these structures).

**TABLE 1.** The RMF (NL3*) results of binding energy (BE), root-mean-square charge radius $r_{ch}$, the quadrupole deformation parameter $\beta_2$ and the energy difference between the ground $(BE)_{g.s}$ and first intrinsic excited state $(BE)_{e.}$ for $^{278-322}118$. (The ratios between the asymmetry parameter for surface $(\eta_s)$ to the center $(\eta_c)$ (i.e., the relative neutron–proton asymmetry of the surface to the center $R_\eta$) are given in the last column. Here, the subscripts $s$ and $c$ stand for the surface and the center region of the nucleus, respectively. The energy is given in MeV.)

| Nucleus | Binding Energy (BE) | Quadrupole Deformation $(\beta_2)$ | Charge Radius $(r_{ch})$ | $\Delta E = (BE)_{g.s.} - (BE)_{e.s.}$ | $R_\eta = \eta_s - \eta_c$ |
|---|---|---|---|---|---|
| $^{278}118$ | 1966.5 | 0.258 | 6.271 | 0.258 | 2.51 |
| $^{280}118$ | 1983.4 | 0.553 | 6.495 | 0.553 | 2.54 |
| $^{282}118$ | 2000.1 | 0.553 | 6.507 | 0.553 | 2.58 |
| $^{284}118$ | 2016.1 | 0.554 | 6.520 | 0.554 | 2.61 |
| $^{286}118$ | 2031.7 | 0.551 | 6.529 | 0.551 | 2.64 |
| $^{288}118$ | 2046.8 | 0.543 | 6.535 | 0.543 | 2.69 |
| $^{290}118$ | 2061.7 | 0.533 | 6.538 | 0.533 | 2.71 |
| $^{292}118$ | 2076.3 | 0.528 | 6.546 | 0.528 | 3.04 |
| $^{294}118$ | 2090.2 | 0.535 | 6.563 | 0.535 | 3.16 |
| $^{296}118$ | 2103.5 | 0.544 | 6.582 | 0.544 | 3.20 |
| $^{298}118$ | 2116.2 | 0.554 | 6.602 | 0.554 | 3.33 |
| $^{300}118$ | 2128.3 | 0.564 | 6.624 | 0.564 | 3.47 |
| $^{302}118$ | 2140.0 | 0.580 | 6.651 | 0.580 | 3.61 |
| $^{304}118$ | 2151.2 | 0.582 | 6.667 | 0.582 | 3.96 |
| $^{306}118$ | 2161.6 | 0.59 | 6.688 | 0.590 | 4.11 |
| $^{308}118$ | 2171.6 | 0.609 | 6.721 | 0.609 | 4.26 |
| $^{310}118$ | 2181.3 | 0.622 | 6.747 | 0.622 | 4.41 |
| $^{312}118$ | 2192.8 | 0.753 | 6.893 | 0.753 | 4.57 |
| $^{314}118$ | 2202.6 | 0.766 | 6.923 | 0.766 | 4.71 |
| $^{316}118$ | 2212.1 | 0.776 | 6.950 | 0.776 | 4.87 |
| $^{318}118$ | 2216.9 | 0.571 | 6.749 | 0.571 | 4.94 |
| $^{320}118$ | 2225.6 | 0.525 | 6.720 | 0.525 | 5.06 |
| $^{322}118$ | 2233.7 | 0.534 | 6.739 | 0.534 | 5.13 |

The detailed analysis of the internal structure of the nucleus can possible from the three dimensional (3D) contour plot of the deformed density. In RMF (NL3*), the densities are obtained only for the positive quadrant of the plane parallel to $z$-axis (the symmetry axis), and are evaluated in the $r_\perp z$ plane, where $x = r_\perp \cos\varphi$ and $y = r_\perp \sin\varphi$ (cylindrical coordinates). The results for positive quadrant are suitably reflected to the other quadrants, giving a complete picture of the nucleus in the $r_\perp z$ plane. The contour plotting of total density (i.e. $\rho_n + \rho_p$) for the ground state of $Z = 118$ and $120$ for $N = 172$, $182$ and $204$ is shown in Fig. 2. The left and right column of the figure is for $^{290,302,322}118$ and $^{292,304,324}120$, respectively. From the figure, one can see the ground states belong to the deformed prolate structure for the isotopic chain of both $Z = 118$ and $120$ (see Tables 1 and 2). The color code along with the density ranges are given in the exact right of the contour plots. From the color code, we can determine the range of the density values for a specific region of the nucleus (i.e. the cluster structures). For example, the color code with 'deep gray' corresponds to maximum density value ($\rho \sim 0.18$ fm$^{-3}$) and the 'light gray' bearing the minimum density ($\rho \sim 0.001$ fm$^{-3}$). A careful inspection of the contour plots of the ground state density distributions shows a broken ring like the structure at the surface of the isotopes along the valley of stability. For example, the appearance of distorted ring structure in case of proton rich isotopes for $N = 172$ ($^{290}118$ and $^{292}120$) are almost disappeared by moving towards the neutron rich isotopes (see Fig. 2). The magnitude of the density in the ring region is higher than that of the central region of a nucleus. It shows a clear signature for some special feature of the nucleus. Beside this, one can observe the neutron skin structure for the neutron rich isotopes of $Z = 118$ and $120$. Hence, the special attribution of density at the surface region of the nucleus plays crucial role in the mode of decay of these nuclei, which will be discussed in the next paragraph extensively.

The asymmetry parameter $\eta$ can be estimate from the mean field density distributions and defined as,

$$\eta = \frac{(\rho_n - \rho_p)}{(\rho_n + \rho_p)}, \tag{4}$$

very high value (i.e. almost four times of the central value) and form a ring-like structure at the surface. Quantitatively, the central region (i.e. $\sim r = 0$ to $5$ fm), with a value of $\eta \sim 0.2$, which drastically changes to $\eta \sim 0.8$ at the surface region (i.e. $\sim r = 5$ to $8$ fm). Here we also got a ring-like structure for all the isotopes (see Fig. 3), but the situation is just reversed of the ground state configuration (see Fig. 2). For example, the distorted ring appears to get into the complete picture with respect to neutron number. More careful inspection of the contour plot shows that the ring structures are also shifted toward the surface with increase of the neutron number. The formation of neutron bunch at the surface region in case of neutron-rich isotopes gives a strong evidence for $\beta^-$ decay. It is worth mentioning that the calculations in triaxially deformed coordinate space may resolve more issues and will throw more light in this direction.

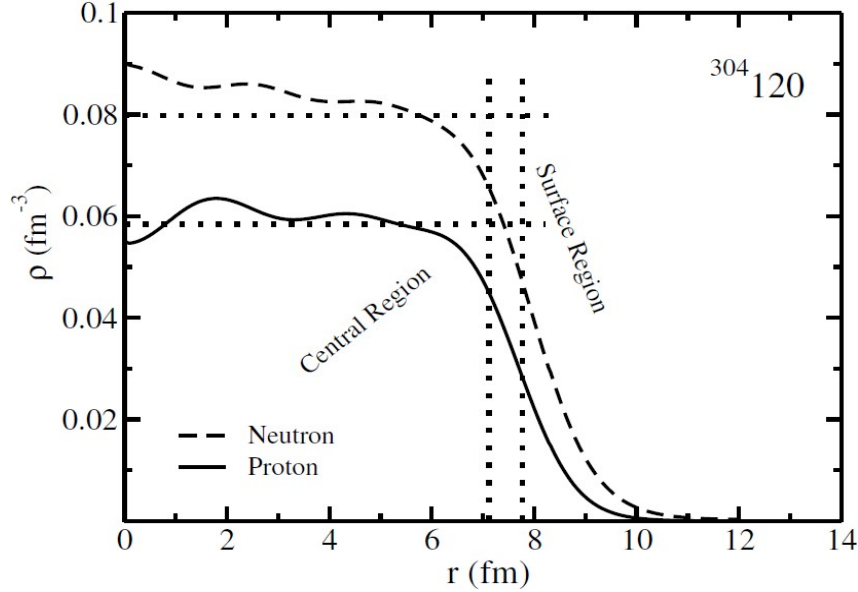

**FIG. 1.** The spherical proton ($\rho_p$) and neutron density ($\rho_n$) distribution for the ground state configuration of $^{304}120$ from the relativistic mean field with NL3* force.

For quantitative analysis, we have estimated the relative neutron–proton asymmetry of the surface to the center of the nucleus $R_\eta = \eta_s/\eta_c$ (i.e. the ratio of the average symmetries of the surface $\eta_s$ to the center $\eta_c$) for the isotopic chains of $Z = 118$ and 120. It is worth mentioning that the ranges for the central region and the surface region are guided by the naked eyes to some extent. As a result, the fraction of the asymmetry parameter $R_\eta$ of the nucleus may be undetermined to the extent of 0.5 units. The estimated relative neutron–proton asymmetry parameters $R_\eta$ are listed in the last column of the Tables 1 and 2 for 118 and 120 isotopes, respectively. From the tables, we found the magnitude of the $R_\eta$ increases with neutron number in a particular isotopic chain. For example, the magnitude of the $R_\eta$ is ∼ 2.5 for $^{290}118$ (i.e. $N = 172$) increases gradually with the neutron number and reach a value of 5.0 for $^{322}118$. We also draw a similar conclusion for the isotopic chain of 120 (see Table 2).

**TABLE 2.** The RMF (NL3*) results of binding energy (BE), root-mean-square charge radius $r_{\text{ch}}$, the quadrupole deformation parameter $\beta_2$ and the energy difference between the ground $(\text{BE})_{\text{g.s}}$ and first intrinsic excited state $(\text{BE})_{\text{e.s}}$ for $^{280-324}$120. (The ratios between the asymmetry parameter for surface $(\eta_s)$ to the center $(\eta_c)$ (i.e., the relative neutron–proton asymmetry of the surface to the center $R_\eta$) are given in the last column. Here, the subscripts $s$ and $c$ stand for the surface and the center region of the nucleus, respectively. The energy is given in MeV.)

| Nucleus | Binding Energy (BE) | Quadrupole Deformation ($\beta_2$) | Charge Radius ($r_{ch}$) | $\Delta\text{E} = (\text{BE})_{\text{g.s.}} - (\text{BE})_{\text{e.s.}}$ | $R_\eta = \eta_s\text{-}\eta_c$ |
|---|---|---|---|---|---|
| $^{280}$120 | 1962.54 | 0.258 | 6.3 | 0.058 | 2.51 |
| $^{282}$120 | 1980.67 | 0.248 | 6.307 | 0.201 | 2.54 |
| $^{284}$120 | 1997.28 | 0.233 | 6.309 | 0.926 | 2.57 |
| $^{286}$120 | 2013.78 | 0.567 | 6.306 | 0.34 | 2.63 |
| $^{288}$120 | 2029.97 | 0.562 | 6.309 | 0.929 | 2.68 |
| $^{290}$120 | 2045.56 | 0.556 | 6.313 | 0.301 | 2.71 |
| $^{292}$120 | 2060.87 | 0.547 | 6.285 | 0.729 | 2.75 |
| $^{294}$120 | 2075.85 | 0.541 | 6.385 | 0.916 | 2.81 |
| $^{296}$120 | 2090.29 | 0.545 | 6.400 | 2.394 | 2.87 |
| $^{298}$120 | 2104.3 | 0.554 | 6.305 | 0.058 | 2.91 |
| $^{300}$120 | 2117.63 | 0.564 | 6.311 | 3.291 | 3.99 |
| $^{302}$120 | 2130.28 | 0.586 | 6.318 | 3.691 | 3.11 |
| $^{304}$120 | 2142.57 | 0.591 | 6.326 | 4.462 | 3.21 |
| $^{306}$120 | 2154.1 | 0.596 | 6.34 | 4.744 | 3.38 |
| $^{308}$120 | 2164.84 | 0.600 | 6.747 | 4.439 | 3.43 |
| $^{310}$120 | 2175.19 | 0.614 | 6.831 | 4.727 | 3.61 |
| $^{312}$120 | 2186.32 | 0.726 | 6.858 | 5.396 | 3.79 |
| $^{314}$120 | 2196.88 | 0.726 | 6.88 | 5.837 | 3.94 |
| $^{316}$120 | 2206.9 | 0.729 | 6.642 | 5.797 | 4.07 |
| $^{318}$120 | 2216.86 | 0.742 | 6.643 | 5.743 | 4.16 |
| $^{320}$120 | 2221.24 | -0.436 | 6.634 | 0.707 | 4.27 |
| $^{322}$120 | 2230.56 | -0.445 | 6.618 | 0.765 | 4.41 |
| $^{324}$120 | 2239.09 | -0.448 | 6.606 | 0.544 | 4.63 |

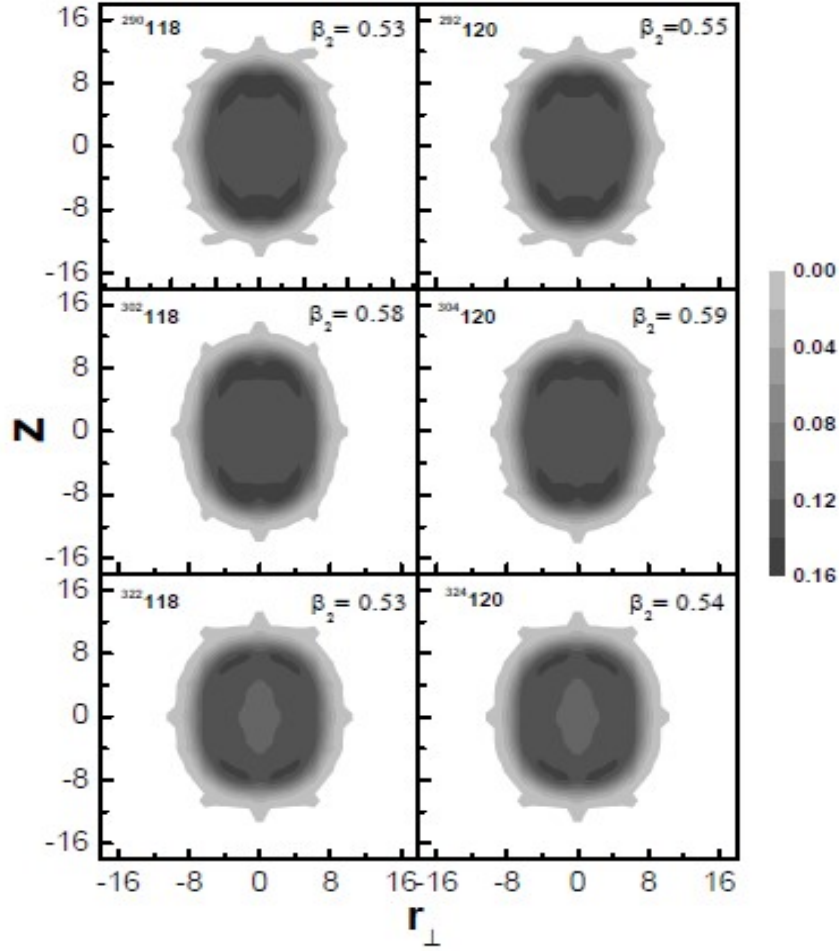

**FIG. 2.** The axially deformed internal ground state configuration of $^{290,302,322}118$ and $^{292,304,324}120$ nuclei from relativistic mean field with NL3* force. The color code along with the density values are displayed in the right side of the figure.

From the present analysis, it is clearly observed that the excess neutron accumulate at the surface region instead of center of the nucleus. By considering the contour plot of the ground state density distribution of the nuclei (i.e., sees Fig. 2), the central density of the nucleus is maximal and it gradually decrease with radius. Unlike the ground state configuration, the magnitude of asymmetry parameter $\eta$ is quite high on the surface than that of center of a nucleus. Similarly, the value of $R_\eta$ gets multiplied in magnitude (i.e., around 3 times) through an isotopic chain. The high values of relative neutron–proton asymmetry at the surface of neutron-rich nuclei are a kind of precursor for $\beta^-$ decay. This effect is also manifested in progressive appearance of the excess neutron, eventually creating a neutron ring-like structure with high neutron density. Due to extreme neutron richness at the surface forms a neutron rings (see Fig. 3), which allowing for the $\beta^-$ decay. Thus the predominant mode of decay of neutron rich superheavy nuclei would $\beta^-$ instead of $\alpha$ decay. The situation seems parallel in the case of a superheavy element in the valley of stability where $\alpha$ decay becomes a more preferred mode of decay over fission. It is also similar kind of interpretation in the fission process of highly neutron-rich nuclei [14, 52], where the fission process is inhibited due to extreme neutron richness in the neck region (i.e., details in [52]).

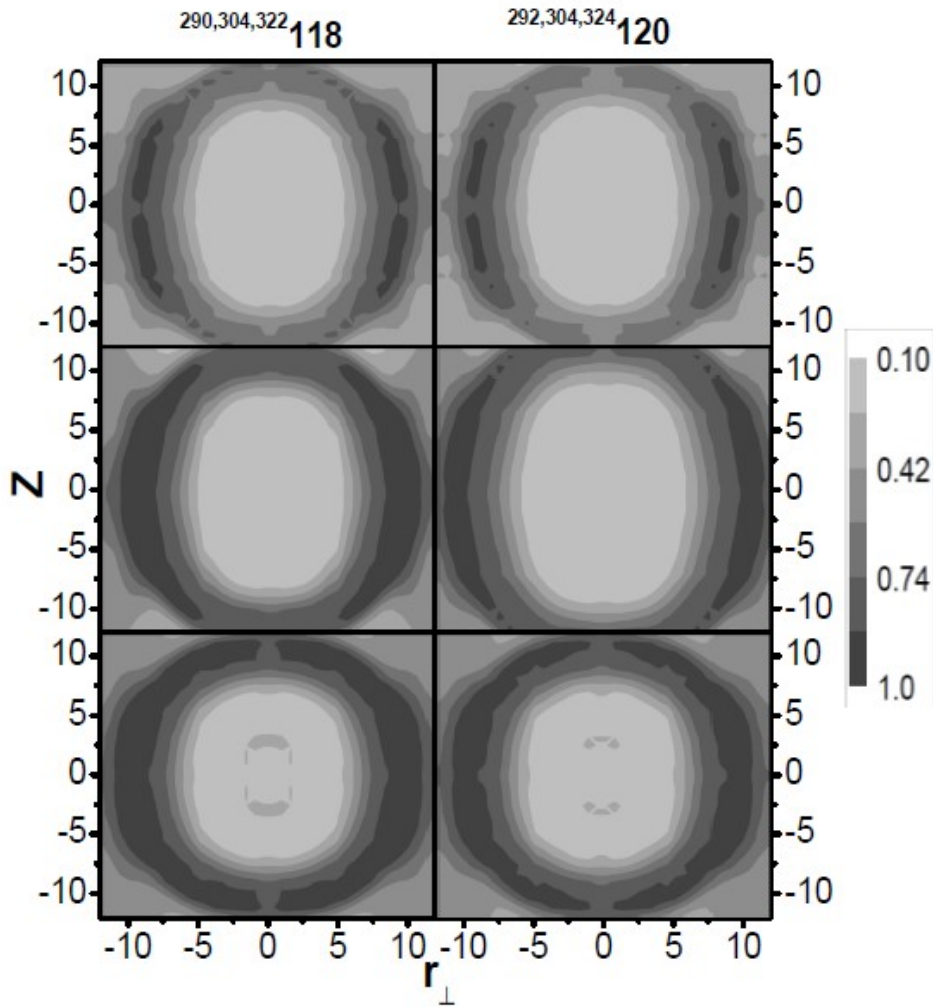

**FIG. 3.** The contour plot of the neutron–proton asymmetry parameter $\eta$ for $^{290,302,322}118$ and $^{292,304,324}120$ nuclei. The color code along with the density values are displayed in the right side of the figure.

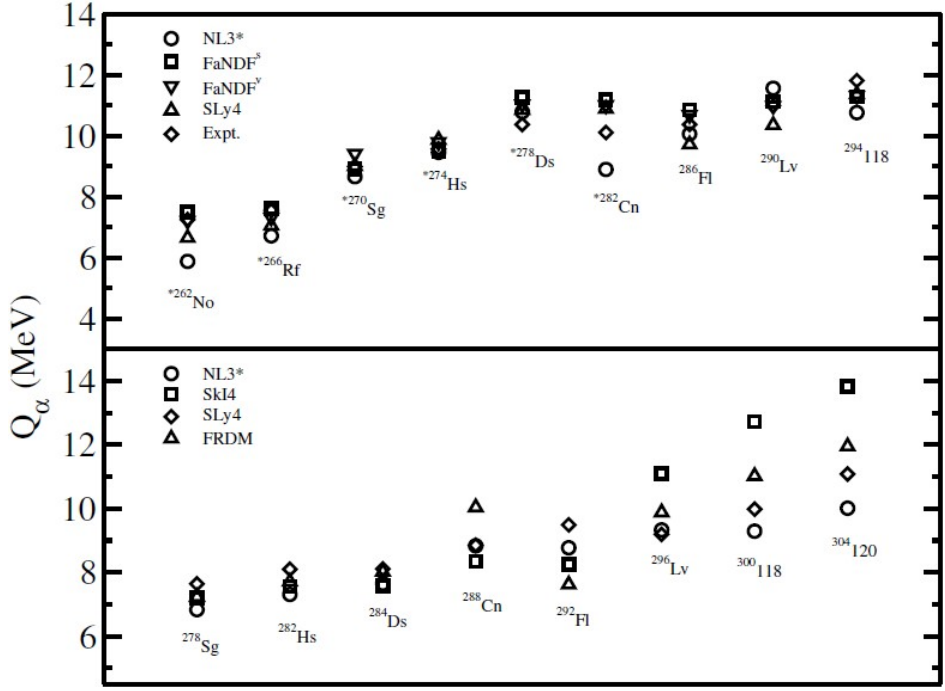

**FIG. 4.** The RMF (NL3*) calculated α decay energy $Q_\alpha$ values for the decay chain of $^{294}118$ and $^{304}120$ are given in the upper and lower panel, respectively. The obtained results are compared with the other theoretical models such as Skyrme-Hartee-Fock with SLy4 [54, 27], SkI4 [27], FaNDF for surface (FaNDF$^s$) and volume (FaNDF$^v$) approximation [54], Finite-Range-Droplet-Model (FRDM) [55, 56], and the experimental data [57, 58], wherever available. The energy is given in MeV.

We know that the α decay process is an essential for the investigation of physical quantities relevant to decay modes of the superheavy nuclei. Further the calculation of the α-decay energy will be informative to confirm the validity of RMF (NL3*) predictions in this regions. Here, the α-decay energy can obtain from the relation [53]: $Q_\alpha$ *(N, Z) = BE (N, Z) − BE (N−2, Z−2) − BE (2, 2)*. Here, *BE (N, Z)* is the binding energy of the parent nucleus with neutron number *N* and proton number *Z*, *BE (2, 2)* is the BE of the α-particle (4He), i.e., 28.296 MeV, and *BE (N−2, Z−2)* is the binding energy of the daughter nucleus after the emission of the α-particle. In the present study, we choose the nucleus $^{294}118$ (*N* = 176) and $^{304}120$ (*N* = 184) for illustrating our calculations for the α-decay chains. The *BE*s of the parent and daughter nuclei are obtained by using the RMF (NL3*) formalism. The calculated $Q_\alpha$ values along with other theoretical models such as Skyrme-Hartee-Fock with SLy [54, 27], SkI4 [27],

FaNDF for surface (FaNDF$^s$) and volume (FaNDF$^v$) approximations [54], Finite-Range-Droplet-Model (FRDM) [55, 56], and the experimental data [57, 58], wherever available, are shown in Fig. 4. The upper and lower panel of the figure is for the decay chains of $^{294}$118 and $^{304}$120, respectively. From the figure, we notice that the calculated values for $Q_\alpha$ coincides quite well with the experimental data and other theoretical models throughout the decay chain of $^{294}$118. Similarly, for $^{304}$120 decay chain, the RMF (NL3*) prediction matches well with all the theoretical predictions of lighter mass nuclei in the decay chain and differ slightly for high mass number. For example, the values of $Q_\alpha$, for RMF (NL3*) coincides well with other theoretical predictions for the chain of $^{296}$Lv, $^{292}$Fl, $^{288}$Cn, $^{284}$Ds, $^{282}$Hs, and $^{278}$Sg, differ slightly for $^{300}$118, and $^{304}$120 (see Fig. 4). This comparison of the $\alpha$-decay energy obtained from the RMF (NL3*) with the experimental data (wherever available) and other theoretical predictions validate the predictive power of the RMF model in this region of the nuclear chart.

## 4. CONCLUSIONS

We have analyzed the ground and first intrinsic excited state bulk properties such as binding energy, charge radius $r_{ch}$, and the binding energy difference of ground and intrinsic first excited states $\Delta E$ for the isotopic chains of $Z = 118$ and 120. The RMF model, which has gained the confidence of the nuclear community in the study of exotic nuclei including superheavy nuclei, has been adopted for the present study. We found a deformed prolate ground state structure of these nuclei followed by a spherical intrinsic excited state. The internal configurations for the ground state solution along with the asymmetry parameter $\eta$ are also estimated. The random distribution of $\eta$, give a primary idea for the $\beta^-$ mode of decay in the neutron-rich superheavy nuclei. Further, the widely varying relative neutron-proton asymmetry of the surface to the center of a nucleus $R_\eta$ strengthens our idea quantitatively for the predominant possible mode $\beta^-$ decay in the neutron rich isotopes of superheavy nuclei. To our knowledge, this is one of the first such interesting and crucial phenomenon presented in the superheavy region. Further we have calculated the $\alpha$ decay energy for the decay chain of $^{294}$118 and $^{304}$120 and compare the results with other theoretical predictions and experimental data, wherever available. We found the RMF (NL3*) results well agreements with other theoretical predictions and experimental data, which shows the validity of the RMF model for the superheavy region of the nuclear chart.

**ACKNOWLEDGMENTS**

This work is supported in part by FAPESP Project No. 2014/26195-5 and CNPq-Brasil. The authors are also thankful to B. V. Carlson, Instituto Tecnológico de Aeronáutica, and S. K. Patra, Institute of Physics, for discussions and support throughout the work.

――――――――――――――――――――――――――――――